\documentclass[showpacs,aps,pra,floatfix,reprint,superscriptaddress,footinbib,citeautoscript]{revtex4-2}
\usepackage{amssymb,amsfonts,amsmath}
\usepackage{graphicx}
\usepackage{verbatim}
\usepackage[utf8]{inputenc}
\usepackage[T1]{fontenc}
\usepackage{soul,xcolor,colortbl}
\usepackage{multirow}
\usepackage{bm}
\usepackage{array}
\usepackage{chemmacros}
\usepackage{textgreek}
\usepackage{color}
\usepackage{hyperref} \hypersetup{colorlinks=true,citecolor=blue,linkcolor=blue,urlcolor=blue}
\usepackage{mdframed}
\usepackage{longtable}
\usepackage{listings}
\usepackage{enumitem}
\usepackage{booktabs}
\usepackage{lipsum}
\usepackage{orcidlink}
\usepackage{siunitx}
\usepackage[normalem]{ulem}

\usepackage{makecell}

\usepackage{xurl}

\newcolumntype{L}[1]{>{\raggedright\arraybackslash}p{#1} }
\newcolumntype{C}[1]{>{\centering \arraybackslash}p{#1} }
\newcolumntype{R}[1]{>{\raggedleft \arraybackslash}p{#1} }

\DeclareSIUnit\inch{inches}
\newcommand{\reffig}[1]{Figure~\ref{#1}}
\newcommand{\newBullet}[0]{$\color{new_entry_green}\mathlarger{\mathlarger{\mathlarger{\mathlarger{\bullet}}}}$}
\newcommand{\oldBullet}[0]{$\mathlarger{\mathlarger{\mathlarger{\mathlarger{\circ}}}}$}
\newlist{noteize}{itemize}{4}
\setlist[noteize]{label=\textcolor{black}{\textbullet}, font=\footnotesize, noitemsep, align=parleft, labelwidth=0.5em, leftmargin=1em}

\def\AFLOW{{\small AFLOW}}

\definecolor{binarycolor}{rgb}{0,0,1}
\definecolor{codegreen}{rgb}{0,0.6,0}
\definecolor{codegray}{rgb}{0.5,0.5,0.5}
\definecolor{codepurple}{rgb}{0.58,0,0.82}
\definecolor{codebgcolorCommand}{rgb}{0.95,0.95,0.92}
\definecolor{codegreen}{rgb}{0,0.6,0}
\definecolor{codegray}{rgb}{0.5,0.5,0.5}
\definecolor{codepurple}{rgb}{0.58,0,0.82}
\definecolor{codebgcolorCommand}{rgb}{0.95,0.95,0.92}
\definecolor{codebgcolorFile}{rgb}{0.92,0.92,0.95}
\definecolor{pranab_green}{rgb}{0.31,0.53,0.10}
\definecolor{pranab_red}{rgb}{0.85,0.23,0.11}
\definecolor{pranab_blue}{rgb}{0,0,0.99}
\definecolor{new_entry_green}{rgb}{0.31,0.8,0.10}

\colorlet{sp_green}{green!65!black}
\colorlet{mp_red}{red!45!black}

\lstdefinelanguage{aflowBash}{
  language=bash,
  basicstyle=\ttfamily\footnotesize,
  morekeywords={aflow,mpirun,vasp46s},
  otherkeywords={install-aflow.sh},
  frame=single,
  breaklines=true,
  backgroundcolor=\color{codebgcolorCommand},
  keywordstyle=\color{binarycolor},
  commentstyle=\it\color{codegreen},
  moredelim=[is][\it]{/*}{*/},
  postbreak=\raisebox{0ex}[0ex][0ex]{\ensuremath{\color{red}\hookrightarrow\space}},
}

\lstdefinelanguage{python}{
  language=python,
  basicstyle=\ttfamily\footnotesize,
  morekeywords={aflow,mpirun,vasp46s},
  otherkeywords={install-aflow.sh},
  frame=single,
  breaklines=true,
  backgroundcolor=\color{codebgcolorCommand},
  keywordstyle=\color{binarycolor},
  commentstyle=\it\color{codegreen},
  moredelim=[is][\it]{/*}{*/},
  postbreak=\raisebox{0ex}[0ex][0ex]{\ensuremath{\color{red}\hookrightarrow\space}},
}

\DeclareSIUnit{\atom}{atom}

\setcitestyle{square}

\makeatletter \renewcommand\frontmatter@abstractwidth{\dimexpr\textwidth\relax} \makeatother

\def\MEMS{\footnotesize Department of Mechanical Engineering and Materials Science, Duke University, Durham, NC 27708, USA}
\def\CEM{Center for Extreme Materials, Duke University, Durham, NC 27708, USA}

\begin{document}
\title{The AFLOW Library of Crystallographic Prototypes: Part 5}
\author{Nicholas~H.~Anderson\,\orcidlink{0000-0000-0000-0000}\affiliation{\MEMS}\affiliation{\CEM}}
\author{Michael~J.~Mehl\,\orcidlink{0000-0001-9402-6591}}\affiliation{\MEMS}\affiliation{\CEM}
\author{Hagen~Eckert\,\orcidlink{0000-0003-4771-1435}}\affiliation{\MEMS}\affiliation{\CEM}
\author{Simon~Divilov\,\orcidlink{0000-0002-4185-6150}}\affiliation{\MEMS}\affiliation{\CEM}
\author{Xiomara~Campilongo\,\orcidlink{0000-0001-6123-8117}}\affiliation{\CEM}
\author{Stefano~Curtarolo\,\orcidlink{0000-0003-0570-8238}}\email[]{stefano@duke.edu}\affiliation{\MEMS}\affiliation{\CEM}
\def\initials{N.A., M.J.M., H.E, S.D., X.C., and S.C.}

\date{\today}

\begin{abstract}
 \noindent
The AFLOW library of crystallographic prototypes has been updated to incorporate an additional 344 entries,
which now reaches 2,127 prototypes. ICSD and CCDC numbers have been added to the website alongside improvements to the user interface. New tutorials covering the basics of crystallography in the context of materials science have also been added. Lastly, we covered the current known applications of AFLOW prototype labels and materials data across research software.
\end{abstract}
\maketitle

\noindent

\section*{Introduction}\label{sec:intro}

The first catalog of minerals dates back to Theophrastus, a student of Aristotle, who compiled a rational classification of stones based on physical properties~\cite{theophrastus_on_stones}.
For another two millennia, descriptions remained qualitative until Steno (1669) and Hauy (1784) hinted at the microscopic ordering of crystal structures through cleavage planes~\cite{Wenk_Bulakh_Minerals}. Hauy's emphasis of crystallographic form drew critics, many supporting Werner, Hauy's contemporary, who had earlier placed chemical composition as the true basis for the division of minerals~\cite{Hazen01111984}. Optical studies of structure-dependent effects (mid 19th century) like birefringence seemed to support Hauy~\cite{Hazen01111984}, encouraging the rigorous enumerations of the crystallographic lattices by Bravais (1850), and space groups by Fedorov (1885-1892) and Schoenflies (1891)~\cite{Wenk_Bulakh_Minerals}, which represented the pinnacle of the theory. By the 20th century, the competing camps of Hauy and Werner were reconciled. The development of X-ray diffraction techniques by Laue (1912)~\cite{eckert2012max}, and further refinement by W.H. and W.L. Bragg (1913)~\cite{Bragg_diamond_1913} confirmed microscopic periodic structure, cementing the modern crystallochemical form of materials identification. What remained was the enormous task of cataloging, something now accomplished easily with modern electronic databases. Across all eras, the mission has been the same: turn chaos into searchable guides.

This has been the spirit infused into the AFLOW library of crystallographic prototypes, started in 2017. Originally developed from NRL's Crystal Lattice Structures (1995), the library has been redesigned to meet the needs of modern materials calculation workflows. The library now provides a holistic crystallographic database suitable for both novices and experts, containing the most commonly encountered structures categorized by a standardized prototype format. Users may visit the encyclopedia for free to browse available prototypes, visualize structures, and generate geometry files for their own calculations.

As we continue to improve and expand our previous work, we document here the current known implementations of the AFLOW prototype label in crystallographic code bases. For a historical overview of crystallographic databases, please see our previous installments in this series of articles~\cite{curtarolo:art121,curtarolo:art145,curtarolo:art173,curtarolo:art210}.

The format of this work is as follows: Section~\ref{sec:updates} discusses the updates to the library, including 344 new structures. Furthermore, this section covers the new video and written tutorials on crystallography in the context of computational physics, including information on chiral space-groups, enantiomorphic pairs, and the AFLOW prototype label itself. Section~\ref{sec:aflow_uses} lists research software that makes either direct use of the AFLOW prototype labels or the AFLOW materials library. The software covered falls under the following broad categories: characterization, database, and visualization codes. The first category concerns programs that extract properties from structural data or are capable of generating new structures altogether. Some examples in this category include crystal identification software like AFLOW Xtalfinder, Spglib, and pymatgen, and structure generators like XtalOPT, and PyXtal. Among electronic databases and visualization software, we cover those focusing on computational workflows as opposed to experimentation.

\section{Updates to the Library}\label{sec:updates}

The library has undergone significant updates since the publication of
Part 4~\cite{curtarolo:art210}.

\subsection{New Prototypes and Updated Prototypes}\label{sec:newstructures}

We have incorporated 344 new prototypes to the encyclopedia, which now includes a total of 2,127.  The additions are listed in the appendix.

Many of these prototypes were chosen to fill ``holes'' in the database, in particular, the sparsity of entries in some space groups.  As an example, space group $P2$ (\#3), which has 69 entries in the current ICSD, only had two entries in the encyclopedia.  We added the monoclinic UMo$_{5}$O$_{16}$~\cite{MonoUMo5O16} and $\alpha$-KHo$_{2}$F$_{7}$~\cite{aKHo2F7} prototypes to help fill in this gap. Others were selected as part of the pedagogical aspect of the encyclopedia.  To highlight the similarities in bonding that can occur between prototypes, many silica (SiO$_{2}$) and aluminum phosphate (AlPO$_{4}$) prototypes were added~\cite{SiO2AlPO4}.

We have also updated existing entries. If a prototype has been entered into the Inorganic Crystal Structure Database (ICSD)~\cite{ICSD,ICSD0,ICSD1,ICSD2,ICSD3,ICSD4,ICSD_database}, we include its database code on the web page. Organic and metal-organic crystals are usually not listed in the ICSD, but their crystallographic information is often entered into the Cambridge Structural Database (CSD)~\cite{Groom_CSD_2016}.  All entries in the ICSD and CSD have a deposition number in the repository maintained by the Cambridge Crystallographic Data Center (CCDC)~\cite{CCDC}.  All ICSD entries have a corresponding entry in the CCDC, so we include both deposition numbers there.  We also include the CCDC deposition number for prototypes which are in the CSD but not in the ICSD.

\subsection{Video Tutorials}\label{sec:videos}

We have added a number of video tutorials to the library.  They describe aspects of crystallography and the relationship of crystal types focusing on applications of these techniques to computational materials physics.  The currently available tutorials include:

\begin{itemize}

\item A description of the AFLOW prototype label~\cite{curtarolo:art121,curtarolo:art210}, including how it is generated and how it can be used to record crystal structure data and as input for AFLOW~\cite{curtarolo:art104} high-throughput electronic structure calculations.

\item Discussion of the distribution of crystal structures in the 230 three-dimensional space groups.  This is broken down into the distribution of structures in the ICSD, the CSD, and by ``structure types,'' our attempt to integrate the ICSD structure types with the AFLOW prototype label.

\item Discussion of chiral space groups, including those which are   enantiomorphic pairs, and those which contain their own mirror images.

\item A series of tutorials on aspects of crystallography applicable to computational physics, including:

\begin{itemize}

\item Defining a lattice and a basis, as well as translational symmetry.

\item A discussion of the seven crystal systems.

\item Distinguishing between conventional and primitive lattices,   including a discussion of the fourteen Bravais lattices.

\item Rotational symmetry in crystals, including the allowed rotations in every crystal system.

\item The crystallographic restriction theorem, which limits the number of rotations allowed in a three-dimensional periodic crystal.

\begin{figure*}
  \centering
  \includegraphics[width=0.30\textwidth]{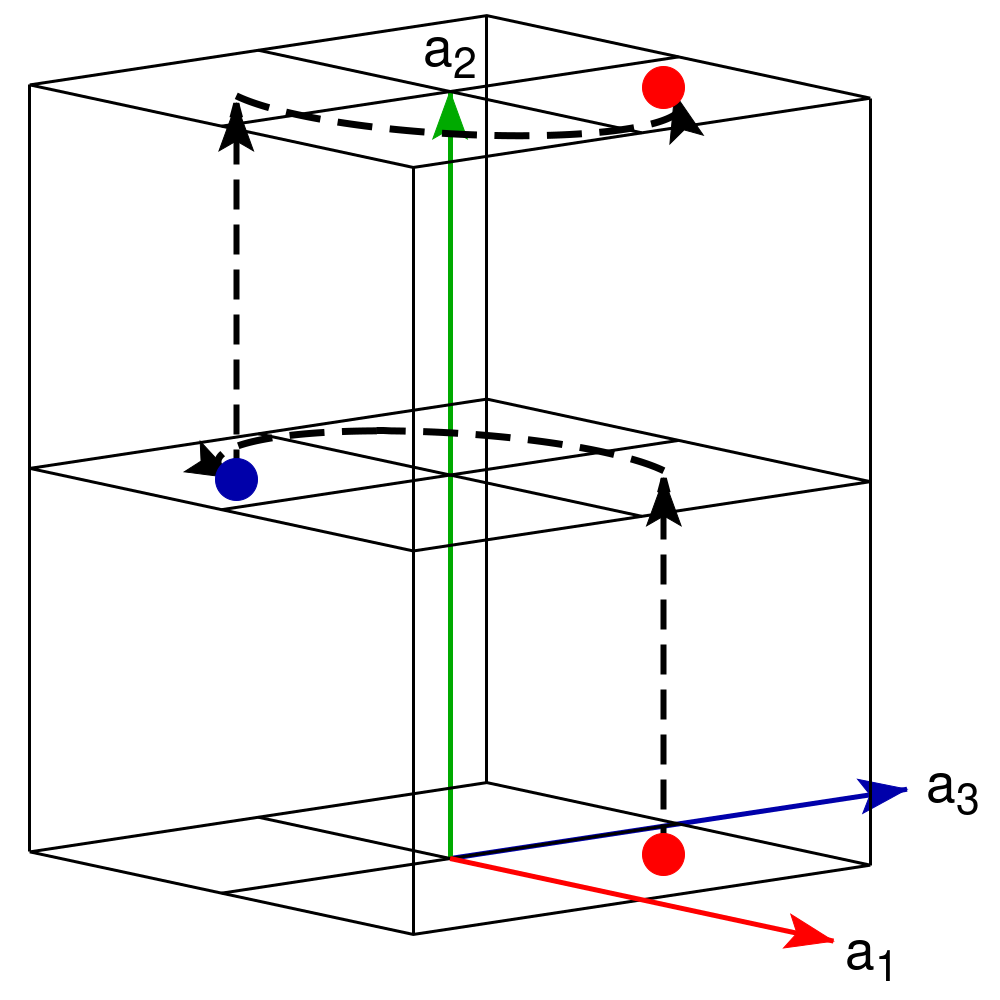}
  \includegraphics[width=0.30\textwidth]{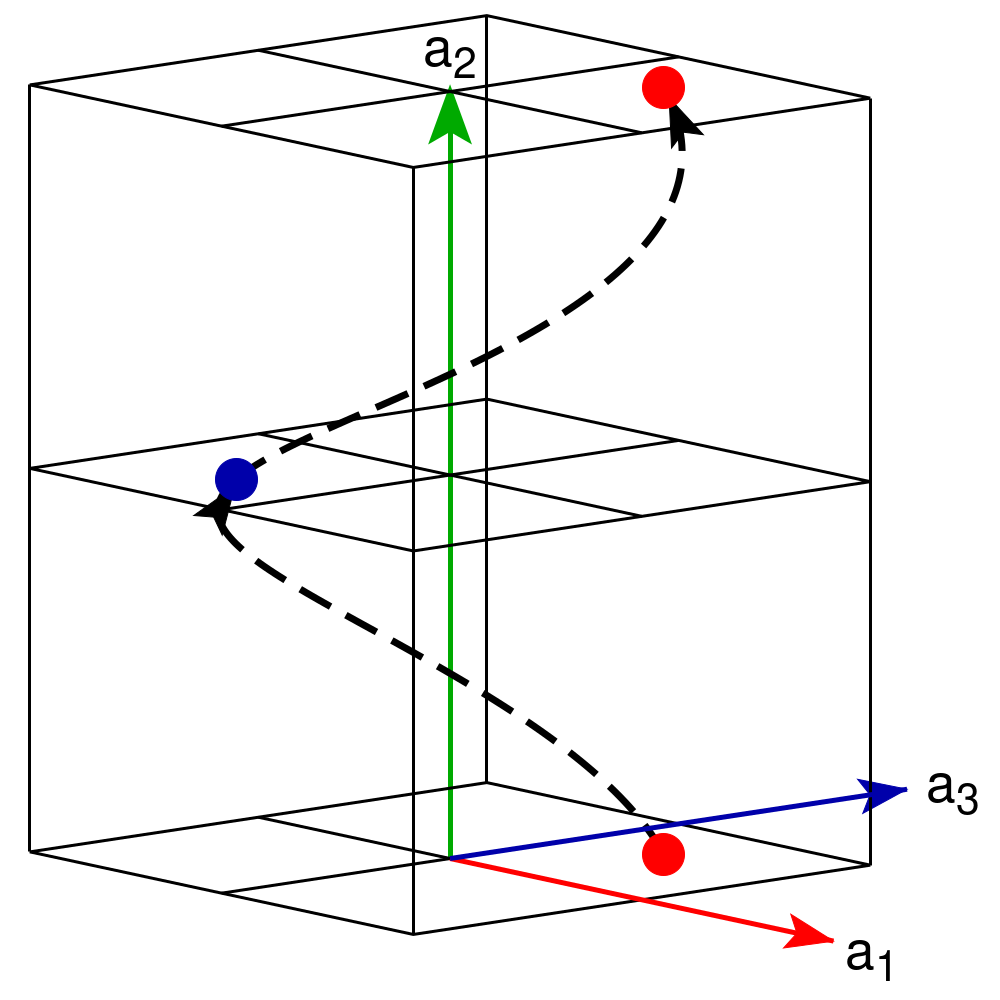}
  \includegraphics[width=0.30\textwidth]{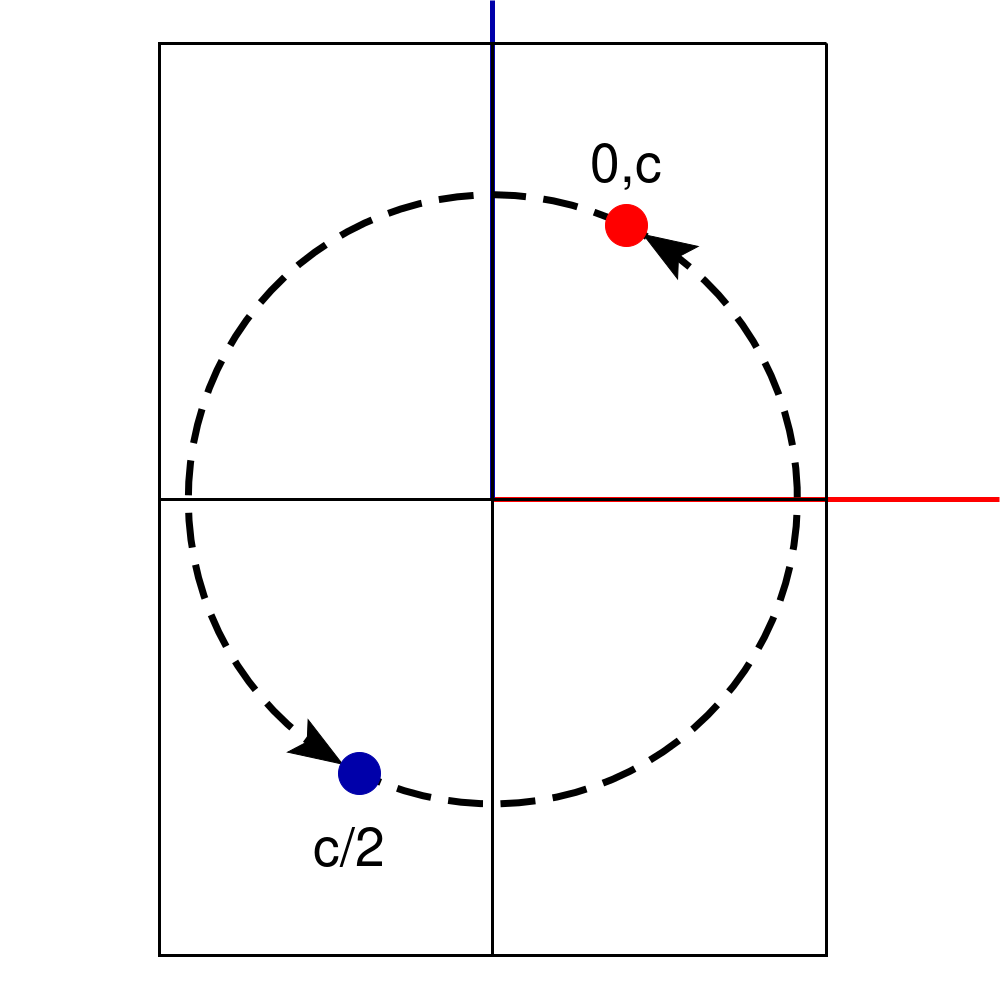}
  \caption{\label{fig:2_1_screw} A diagram of the operations needed to construct a $2_{1}$ screw axis in an orthorhombic crystal. The ``atom'' colors are selected to show the different heights of the atoms in the unit cell.  In reality all of the atoms in a given screw are identical. Left: $c/2$ translation first, followed by a 180$^\circ$ rotation. Center: the translation and rotation combined, showing the screw axis. Right: A top view.  The labels on each atom show its height above the bottom of the unit cell.}
\end{figure*}

\item Screw axes, viewed as a combination of allowed translations and   rotations.  We show an example of the operations needed for a $2_{1}$ screw in~\reffig{fig:2_1_screw}.

\item The special case of 3-fold screw axes in rhombohedral and cubic crystals.

\end{itemize}

\end{itemize}

All tutorials are available under the {\tt Tutorials} tab on the encyclopedia home page~\cite{NoteTutorials}.

\subsection{Written Tutorials}\label{sec:writings}

Each video tutorial described above is accompanied by a written tutorial.  They expand upon the topics in the videos.

In addition, we have several written tutorials which do not include a video component. These include:
\begin{itemize}

\item A discussion of crystallography in two dimensions, including lattices, basis, translational symmetry, other symmetries, allowed lattices, and a complete description of the seventeen allowed plane groups in two dimensions.  This can serve as an introduction to crystallographic concepts which are easier to visualize on a screen.

\item A discussion of the many possible crystal structures available using SiO$_{2}$ and AlPO$_{4}$ as building blocks, including quartz-like structures as well as open structures such as zeolites.

\end{itemize}

Future written tutorials will include a listing of the structures which can be derived from perovskite, including the high-temperature cuprate superconductors.

In addition to the written tutorials, individual web pages discuss relationships between structures. For instance, the quaternary Heusler (LiMgAuSn)~\cite{LiMgAuSn} page contains a list of the many cubic structures that can be constructed by changing the chemical composition of the structure or removing atoms.
Temperature-driven phase transitions in NaH$_{4}$NO$_{3}$~\cite{NaH4NO3} are also explored.

As with the video tutorials, the stand-alone written tutorials  are available under the ``Tutorials'' tab on the encyclopedia home page~\cite{NoteTutorials}.

\begin{figure*}
  \centering
  \includegraphics[width=0.9\textwidth]{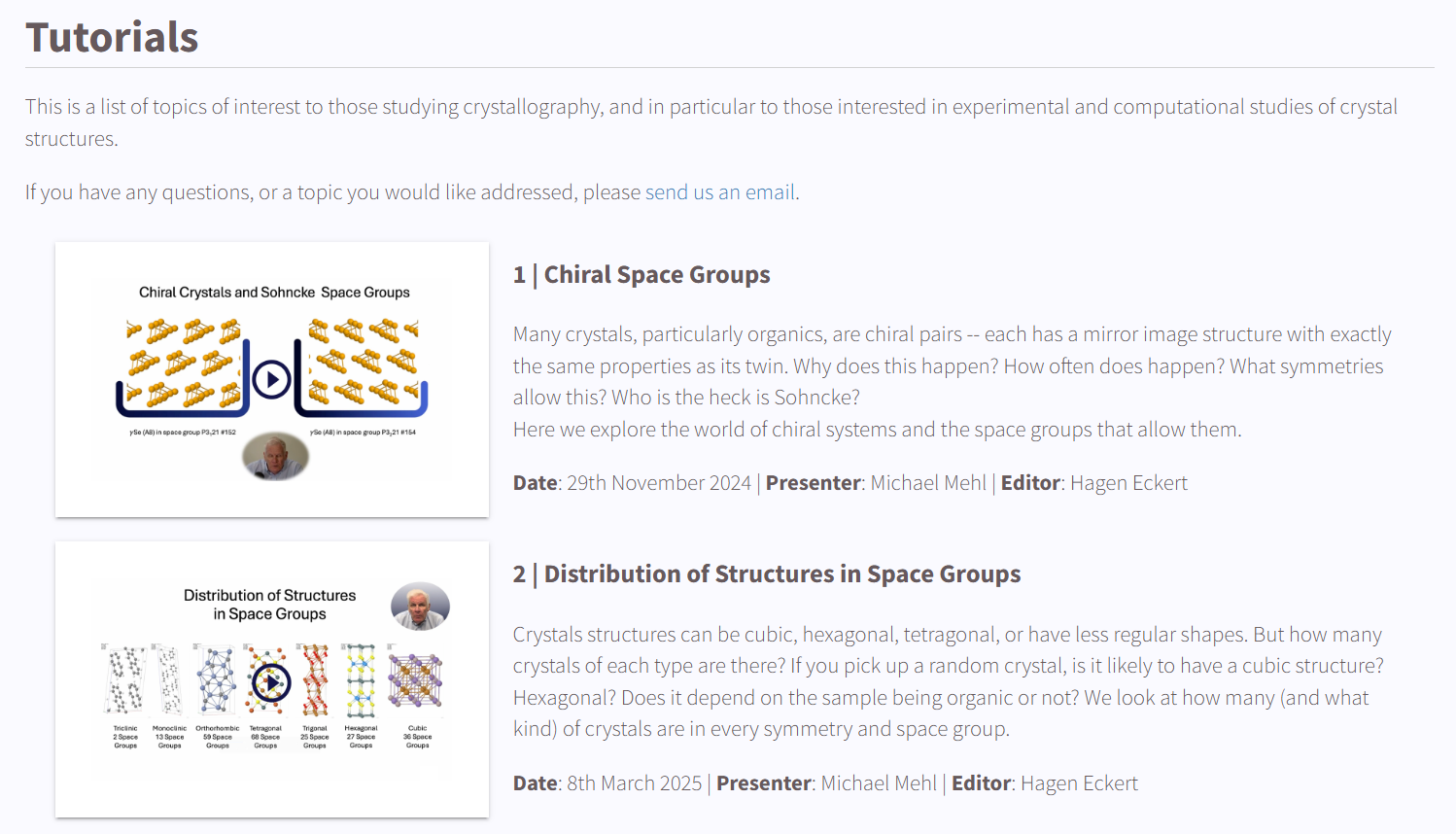}
  \caption{\label{fig:tutorial_page} \texttt{aflow.org} tutorial page available in the prototype encyclopedia.}
\end{figure*}

\vspace{1cm}

\section{Software That Use Aflow Prototypes}\label{sec:aflow_uses}

\begin{table*}
\centering
 \caption{List of active research software and databases that use AFLOW prototypes or \texttt{aflow.org} data in a crystallographic context.}
 \setlength{\tabcolsep}{12pt}
\begin{tabular}{l l l l}
 \hline
 Software & Source(s) & Language(s)  & Description\\
 \hline
 OpenKIM &~\cite{OpenKIM}& C++, Fortran & Interatomic potential database\\
 kim-tools &~\cite{kim-tools}& python & Creation of new OpenKIM test drivers \\
 NOMAD &~\cite{NOMAD_2023}& python, JavaScript & Materials database \\
 OPTIMADE API &~\cite{curtarolo:art208_UPDATED}& Make, python & Inter-database API \\
 pymatgen &~\cite{pymatgen}& python & Robust library for materials analysis\\
 robocrystallographer &~\cite{Robocrystallographer, robocrystallographer_source} & python  & Symmetry-based text descriptions of crystals \\
 {Jmol} &~\cite{Jmol,Jmol_Hanson}& {Java} & {Crystal and molecule visualization} \\
 {{aviary}} &~\cite{aviary, riebesell_2025_framework}& {{python}} & {{Multi-model interface for materials discovery}} \\
 {{WyckoffDiff}} &~\cite{WyckoffDiff_source, pmlr-v267-ekstrom-kelvinius25a}& {{python}} & {{Generative model for crystal symmetry}} \\
 Wyckoff Transformer &~\cite{Kazeev202529495, Wyckoff_Transformer_source}& python & Symmetric crystal generative model \\
 ARISE &~\cite{ARISE_2021, ARISE_source}& python & ML-based crystal-structure identification \\
 simmate &~\cite{simmate}& python & Chemistry research,  materials analysis \\
 \hline
   \end{tabular}
\end{table*}

Since the publication of the first AFLOW prototypes paper in 2017~\cite{curtarolo:art121}, the greater materials science community has had several years to explore the prototyping library. We cover here the current known implementations from online databases, API's, and crystallographic tool sets where we briefly delve into the features of each. We found additional usages of \texttt{aflow.org} data that we have also indexed in this section.

The most visible usages of the AFLOW prototypes are on the online databases. Among them are the Open Knowledgebase of Interatomic Models (OpenKIM) and the Novel Materials Discovery Lab (NOMAD). OpenKIM provides a standardized approach for validating the accuracy of inter-atomic models by using reference data from an online database of first-principles calculations and experimental results~\cite{OpenKIM}. Included in the calculated reference data are datasets from \texttt{aflow.org}, Materials Project, and individual contributions. A web front-end allows users to browse materials categorized by their AFLOW prototype label. Users can then navigate to information on the compatible interatomic potentials, visualizations, and crystallographic information. Furthermore, OpenKIM hosts a variety of post-processing pipelines including the ``Crystal Genome'' framework for calculating both static and dynamic materials properties. The framework enforces model integrity, requiring each model to pass through a series of verification checks and a set of material property computations which can be examined for accuracy relative to reference data. Users may design their own testing frameworks by using either the Docker image KIM Developer Platform or the standalone kim-tools python package. The Crystal Genome framework classifies crystals with AFLOW prototype labels, requiring an AFLOW installation for users choosing the standalone kim-tools package. The interatomic models developed on OpenKIM have seen use in popular molecular dynamics software such as LAMMPS, ASE, MDStressLab, and by organizations such as NIST.

NOMAD is a web-based application that provides a common framework for the storage and retrieval of materials research data among disparate databases~\cite{NOMAD_2023}. The current web database includes theoretical contributions from \texttt{aflow.org}, Materials Project, and OQMD. Individuals may upload calculations after creating an account. A detailed search system allows specification of the calculation type, workflow, properties, and structural information including use of the AFLOW prototypes as a search field. Information retrieval is straightforward; users do not need to know the data storage methodologies of each database and can query for entries in a standardized format. NOMAD conforms to FAIR standards (Findable, Accessible, Interoperable, Reusable)~\cite{Wilkinson_FAIR_SciData_2016_UPDATED}, implementing the OPTIMADE API along with NOMAD specific APIs for ease of data submission and reusability. NOMAD Oasis, a local version of the web-based NOMAD, has seen adoption by several premier institutes and universities including the Helmholtz Institute, the Leibniz Institute for Crystal Growth, and the University of Illinois, among others ~\cite{NOMAD_about}.

The Open Databases Integration for Materials Design (OPTIMADE) API defines a common interface for the retrieval of materials data, improving the discoverability of data, especially from smaller, lesser-known databases~\cite{curtarolo:art208_UPDATED}. Aligned also with FAIR standards, OPTIMADE retrieves data in a standardized JSON format inspired by CIF. We host an OPTIMADE API that is built off of AFLUX to offer a common search syntax across multiple materials databases. AFLOW property labels are mapped to those of OPTIMADE while other AFLOW-specific properties can be retrieved with the \texttt{\_aflow\_} search prefix. As of 2026, the OPTIMADE API format has been adopted by 28 known providers, including \texttt{aflow.org}, representing 45 sub-databases~\cite{OPTIMADE_about}.

In addition to their usage in web databases, the AFLOW prototype labels have also been used internally in crystallographic programs to assist in structure identification. AFLOW's XtalFinder was the first to pair an internal symmetry routine (AFLOW-SYM) with prototype identification, addressing issues like compound duplication and enabling efficient mappings of structure-property relations of materials~\cite{aflowXTALFINDER}. As a part of its robust materials analysis tool set, the open-source python library pymatgen contains the AflowPrototype-Matcher class that returns the AFLOW prototype label of a given structure using data from the AFLOW prototype library~\cite{CMS_Ong2012b}. Structures in pymatgen are handled internally as a class that can be instantiated with information from standardized crystallographic formats like CIF. Like XtalFinder, pymatgen compares structures within tunable tolerances; however, only XtalFinder is able to independently identify crystallographic symmetries~\cite{aflowXTALFINDER}.

The open-source toolkit robocrystallographer builds on pymatgen's AflowPrototype-Matcher class to provide analysis of the semi-local crystallographic environment using the AFLOW prototype library for framework analysis~\cite{Robocrystallographer}. Properties such as the local coordination, polyhedral type, and octahedral tilt angles are extracted and presented in JSON format as well as in human-readable text descriptions. This code base has the potential to be used in statistical machine learning (ML). As a demonstration, the elastic moduli of a trial set of 1181 inorganic compounds were found to be better reproduced with semi-local properties as opposed to just compositional information alone. In addition, the human-readable text descriptions could be added to databases like Materials Project to enhance materials descriptions and to provide a debugging tool.

Jmol is an open-source Java/SwingJS application for the visualization and  analysis of 3D molecular structures~\cite{Jmol,Jmol_Hanson}. Jmol has a rich scripting language. Users can visualize structures derived from an AFLOW prototype label. For example:

\begin{lstlisting}[language=aflowBash]
load=aflowlib/AB_cF8_225_a_b-001 packed
\end{lstlisting}

\noindent
 displays the sodium chloride structure. In addition, Jmol can also display AFLOW prototypes for use as examples of crystal structures from any of the 230 space group types. For example:

\begin{lstlisting}[language=aflowBash]
load=aflowlib/125.1 packed
\end{lstlisting}

 \noindent
  displays the first representative of space group $P4/nbm$ \#125 found in the encyclopedia. In addition, Jmol has an extensive ``model kit'' that can be used to construct crystal structures de novo for any space group or by symmetry-aware adjustment of atom positions or addition of atoms starting with AFLOW prototypes. Jmol's crystal structure visualization tools are extensive, including the selective visualization of symmetry elements.

As a complement to the larger projects, we were also able to find a host of smaller, more specialized projects that utilized either the prototyping library or \texttt{aflow.org} data. In line with the growing popularity of ML there exist several code bases that calculate novel crystal structures, potentials, and material properties from training data. Aviary is a multi-model interface for materials discovery containing the following routines: Roost, Wren, Wrenformer, and Cgcnn. Wren (Wyckoff REpresentation regressioN) performs machine learning on a set of prototypes, taking their Wyckoff representations as principal input to predict the formation energies of inorganic crystals~\cite{goodall_2022_rapid}. AFLOW prototype labels are used internally to store and parse structural information. In Wren's proof-of-concept studies, the labels were used to reduce the prevalence of low-symmetry and high energy structures in training data. Labels were also used to confirm the novelty of output structures, identifying isopointal prototypes that did not exist in the training data. Wren demonstrated an advantage over substitution-based algorithms, identifying stable structures with 5 times lower computational effort. A more recent variant of Wren, Wrenformer, boasts performance improvements and support for structures with more than 16 Wyckoff positions~\cite{riebesell_2025_framework}.

WyckoffDiff is a symmetry-based generator for novel crystal structures~\cite{pmlr-v267-ekstrom-kelvinius25a}. Similar in approach to Wren, WyckoffDiff is a diffusion model that enforces knowledge of the Wyckoff positions during calculation, efficiently searching the most promising candidates with high symmetry. WyckoffDiff adopts the AFLOW prototype labeling scheme to represent structures internally. Compared to random substitutional generative models like CDVAE, and competing symmetry-inclusive models like DiffCSP++ and SymmCD, WyckoffDiff demonstrated a computational advantage in the number of novel structures generated per minute.

Wyckoff Transformer is an open-source python library that implements ML methods conditioned on the space group, Wyckoff positions, and elements present, to predict novel stable structures and calculate material properties such as thermal conductivity and bulk and shear moduli~\cite{Kazeev202529495}. The \texttt{aflow.org} database was used as part of a training set for the material properties calculations which compared favorably to models that used full structures.

The open-source python library ARISE identifies crystal structures from bulk, 1D, and 2D materials using a deep learning approach designed to correctly characterize strongly perturbed single and polycrystalline systems~\cite{ARISE_2021}. Training datasets were compiled in part from the AFLOW library of crystallographic prototypes. Given a geometry file, ARISE can return the classification probability for an AFLOW prototype label.

Simmate is a python-based framework for easy exploration of third-party databases and calculation of material properties with a chemistry focus~\cite{simmate}. Users can run from a set of preconfigured workflows including materials relaxation, static-energy, and electronic-structure, among others. Included is an AFLOW ``app'' which allows easy installation of the AFLOW prototypes library.

\section*{Conclusion}\label{sec:conclusion}

This article captures the most recent additions to the AFLOW library of crystallographic prototypes including 344 new structural entries and documentation on legacy prototypes. Navigation on the website has been improved by adding both ICSD and CCDC numbers as search fields.
The new video and written tutorials cover the basics of crystallography, computational materials science, symmetry groups and operations, and how to use AFLOW prototype labels. We list the current known implementations of the AFLOW prototypes and \texttt{aflow.org} data in research software. There exist wide-ranging applications from large online materials databases to smaller, specialized repositories for identifying structural and materials properties.

\section{Software License}

We welcome the use of the AFLOW prototype encyclopedia which is free and publicly  available on GitHub under the Apache License version 2.0.

\section*{Declaration of Competing Interest}
The authors declare that they have no competing financial interests
or personal relationships that could have appeared to influence the work
reported in this article.

\section*{Acknowledgments}
The authors thank {David Hicks, Harold Stokes, Paolo De Angelis, Scott Thiel, and Robert Hanson} for fruitful discussions.  As always, the staff of the Duke University Libraries were extremely helpful in obtaining hard-to-find digitalized publications.
This research was supported by the Office of Naval Research under grant {N00014-24-1-2502}.
This work was supported by high-performance computer time and resources from the DoD High Performance Computing Modernization Program (Frontier).
We also acknowledge Auro Scientific, LLC for computational support.
\\

\section*{Author Contributions}
\textbf{Nicholas H. Anderson}: Writing – original draft, software search. \textbf{Michael J. Mehl}:
Writing – review \& editing, Writing – original draft, Visualization,
Formal analysis, Conceptualization. \textbf{Hagen Eckert}: Writing – original draft, Visualization.
\textbf{Simon Divilov}: Writing – software.
\textbf{Xiomara
Campilongo}: Writing – review \& editing, Writing – original
draft. \textbf{Stefano Curtarolo}: Writing – review \& editing, Writing –
original draft, Supervision, Software, Project administration, Funding
acquisition, Conceptualization.

\section*{Data availability}
The prototype data is freely available from the \url{aflow.org} webpage. A structured collection is available at \url{https://github.com/aflow-org/aflow_prototype_encyclopedia}.

\section*{Code availability}
The underlying software used in this study and its source code can be accessed via this link: \url{https://aflow.org/install-aflow/}.

\appendix

\section{Index of Prototypes Ordered by Space Groups}

This index lists all of the prototype structures in the current
encyclopedia of crystallographic prototypes.  Entries headed by a green
bullet ($\color{new_entry_green}\mathlarger{\mathlarger{\bullet}}$) are new to this edition of the encyclopedia.
Clicking on the hash or {\AFLOW} prototype label will open a web
browser window showing the details of that structure. A PDF version of
the page can be obtained by clicking the ``PDF version'' link on the
web page.

\begin{longtable}{lll}
\hline
\multicolumn{3}{l}{$\bf P1 ~ (1)$:} \\
\newBullet & \multicolumn{2}{l}{\makecell[lt]{ AlO$_{4}$P:}} \\
& \href{https://aflow.org/p/AB4C_aP240_1_40a_160a_40a-001}
{\makecell[lt]{\tt AB4C\_aP240\_1\_40a\_160a\_40a-001}}
& \href{https://aflow.org/p/FWJQ}{\tt FWJQ}
\\
\oldBullet & \multicolumn{2}{l}{\makecell[lt]{ AsKSe$_{2}$:}} \\
& \href{https://aflow.org/p/ABC2_aP16_1_4a_4a_8a-001}
{\makecell[lt]{\tt ABC2\_aP16\_1\_4a\_4a\_8a-001}}
& \href{https://aflow.org/p/GZNE}{\tt GZNE}
\\
\oldBullet & \multicolumn{2}{l}{\makecell[lt]{ FeS$_{2}$:}} \\
& \href{https://aflow.org/p/AB2_aP12_1_4a_8a-001}
{\makecell[lt]{\tt AB2\_aP12\_1\_4a\_8a-001}}
& \href{https://aflow.org/p/RHYA}{\tt RHYA}
\\
\oldBullet & \multicolumn{2}{l}{\makecell[lt]{ C$_{5}$H$_{11}$NaO$_{8}$S:}} \\
\end{longtable}

\newcommand{\Ozolins}{Ozoli{\c{n}}{\v{s}}}


\begin{thebibliography}{10}
\expandafter\ifx\csname urlstyle\endcsname\relax
  \providecommand{\doi}[1]{doi:\discretionary{}{}{}#1}\else
  \providecommand{\doi}{doi:\discretionary{}{}{}\begingroup
  \urlstyle{rm}\Url}\fi
\providecommand{\selectlanguage}[1]{\relax}
\providecommand{\bibAnnoteFile}[1]{%
  \IfFileExists{#1}{\begin{quotation}\noindent\textsc{Key:} #1\\
  \textsc{Annotation:}\ \input{#1}\end{quotation}}{}}
\providecommand{\bibAnnote}[2]{%
  \begin{quotation}\noindent\textsc{Key:} #1\\
  \textsc{Annotation:}\ #2\end{quotation}}

\bibitem{theophrastus_on_stones}
E.~R. Caley and J.~F.~C. Richards, \emph{Theophrastus On Stones} (The Ohio
  State University Press, 1956).
\input{theophrastus_on_stones}

\bibitem{Wenk_Bulakh_Minerals}
H.~R. Wenk and A.~Bulakh, \emph{Minerals: Their Constitution and Origin}
  (Cambridge University Press, 2016).
\input{Wenk_Bulakh_Minerals}

\bibitem{Hazen01111984}
R.~M. Hazen, \emph{Mineralogy: A Historical Review}, Journal of Geological
  Education \textbf{32}, 288--298 (1984), \doi{10.5408/0022-1368-32.5.288}.
\input{Hazen01111984}

\bibitem{eckert2012max}
M.~Eckert, \emph{Max von Laue and the discovery of X-ray diffraction in 1912},
  Annalen der Physik \textbf{524}, A83--A85 (2012),
  \doi{10.1002/andp.201200724}.
\input{eckert2012max}

\bibitem{Bragg_diamond_1913}
W.~H. Bragg and W.~L. Bragg, \emph{The structure of the diamond}, Proceedings
  of the Royal Society of London. Series A, Containing Papers of a Mathematical
  and Physical Character \textbf{89}, 277--291 (1913),
  \doi{10.1098/rspa.1913.0084}.
\input{Bragg_diamond_1913}

\bibitem{curtarolo:art121}
M.~J. Mehl, D.~Hicks, C.~Toher, O.~Levy, R.~M. Hanson, G.~L.~W. Hart, and
  S.~Curtarolo, \emph{The {AFLOW} Library of Crystallographic Prototypes: Part
  1}, Comput.\ Mater.\ Sci. \textbf{136}, S1--S828 (2017),
  \doi{10.1016/j.commatsci.2017.01.017}.
\input{curtarolo:art121}

\bibitem{curtarolo:art145}
D.~Hicks, M.~J. Mehl, E.~Gossett, C.~Toher, O.~Levy, R.~M. Hanson, G.~L.~W.
  Hart, and S.~Curtarolo, \emph{The {AFLOW} Library of Crystallographic
  Prototypes: Part 2}, Comput.\ Mater.\ Sci. \textbf{161}, S1--S1011 (2019),
  \doi{10.1016/j.commatsci.2018.10.043}.
\input{curtarolo:art145}

\bibitem{curtarolo:art173}
D.~Hicks, M.~J. Mehl, M.~Esters, C.~Oses, O.~Levy, G.~L.~W. Hart, C.~Toher, and
  S.~Curtarolo, \emph{The {AFLOW} Library of Crystallographic Prototypes: Part
  3}, Comput.\ Mater.\ Sci. \textbf{199}, 110450 (2021),
  \doi{10.1016/j.commatsci.2021.110450}.
\input{curtarolo:art173}

\bibitem{curtarolo:art210}
H.~Eckert, S.~Divilov, M.~J. Mehl, D.~Hicks, A.~C. Zettel, M.~Esters,
  X.~Campilongo, and S.~Curtarolo, \emph{The AFLOW Library of Crystallographic
  Prototypes: Part 4}, Comput.\ Mater.\ Sci. \textbf{240}, 112988 (2024),
  \doi{10.1016/j.commatsci.2024.112988}.
\input{curtarolo:art210}

\bibitem{MonoUMo5O16}
\emph{Monoclinic UMo$_{5}$O$_{16}$ Structure}. \url{https://aflow.org/p/LMET}.
\input{MonoUMo5O16}

\bibitem{aKHo2F7}
\emph{$\alpha$-KHo$_{2}$F$_{7}$ Structure}. \url{https://aflow.org/p/C4TT}.
\input{aKHo2F7}

\bibitem{SiO2AlPO4}
\emph{Silica (SiO$_{2}$) and Aluminum Phosphate (AlPO$_{4}$) Structures
  Tutorial}. \url{https://aflow.org/p/Tutorials/SiO2_AlPO4}.
\input{SiO2AlPO4}

\bibitem{ICSD}
G.~Bergerhoff, R.~Hundt, R.~Sievers, and I.~D. Brown, \emph{The inorganic
  crystal structure data base}, J.\ Chem.\ Inf.\ Comput.\ Sci. \textbf{23},
  66--69 (1983), \doi{10.1021/ci00038a003}.
\input{ICSD}

\bibitem{ICSD0}
A.~D. Mighell and V.~L. Karen, \emph{NIST Materials Science Databases}, Acta\
  Crystallogr.\ Sect.\ A \textbf{49}, c409 (1993),
  \doi{10.1107/S0108767378088492}.
\input{ICSD0}

\bibitem{ICSD1}
V.~L. Karen and M.~Hellenbrandt, \emph{Inorganic crystal structure database:
  new developments}, Acta\ Cryst. \textbf{A58}, c367 (2002).
\input{ICSD1}

\bibitem{ICSD2}
I.~D. Brown, S.~C. Abrahams, M.~Berndt, J.~Faber, V.~L. Karen, W.~D.~S.
  Motherwell, P.~Villars, J.~D. Westbrook, and B.~McMahon, \emph{Report of the
  Working Group on Crystal Phase Identifiers}, Acta\ Cryst. \textbf{A61},
  575--580 (2005).
\input{ICSD2}

\bibitem{ICSD3}
A.~Belsky, M.~Hellenbrandt, V.~L. Karen, and P.~Luksch, \emph{New developments
  in the {In}organic {C}rystal {S}tructure {D}atabase ({ICSD}): accessibility
  in support of materials research and design}, Acta\ Crystallogr.\ Sect.\ B
  \textbf{58}, 364--369 (2002), \doi{10.1107/S0108768102006948}.
\input{ICSD3}

\bibitem{ICSD4}
{FIZ~Karlsruhe~and~NIST}, \emph{Inorganic Crystal Structure Database},
  http://icsd.fiz-karlsruhe.de/.
\input{ICSD4}

\bibitem{ICSD_database}
{FIZ~Karlsruhe}, \emph{Inorganic Crystal Structure Database},
  http://icsd.fiz-karlsruhe.de/ (1998).
\input{ICSD_database}

\bibitem{Groom_CSD_2016}
C.~R. Groom, I.~J. Bruno, M.~P. Lightfoot, and S.~C. Ward, \emph{The Cambridge
  Structural Database}, Acta\ Crystallogr.\ Sect.\ B \textbf{72}, 171--179
  (2016), \doi{10.1107/S2052520616003954}.
\input{Groom_CSD_2016}

\bibitem{CCDC}
\emph{Cambridge Crystallographic Data Center}. Individual structures may be
  found using the ICSD or CCDC deposition number or information from the
  original publication at \url{https://www.ccdc.cam.ac.uk/structures/}.
\input{CCDC}

\bibitem{curtarolo:art104}
C.~E. Calderon, J.~J. Plata, C.~Toher, C.~Oses, O.~Levy, M.~Fornari, A.~Natan,
  M.~J. Mehl, G.~L.~W. Hart, M.~{Buongiorno Nardelli}, and S.~Curtarolo,
  \emph{The {AFLOW} standard for high-throughput materials science
  calculations}, Comput.\ Mater.\ Sci. \textbf{108}, 233--238 (2015),
  \doi{10.1016/j.commatsci.2015.07.019}.
\input{curtarolo:art104}

\bibitem{NoteTutorials}
\protect \url {https://aflow.org/prototype-encyclopedia/tutorials.html}.
\input{NoteTutorials}

\bibitem{LiMgAuSn}
\emph{Quaternary Heusler (LiMgAuSn) Structure}.
  \url{https://aflow.org/p/4CEX/}.
\input{LiMgAuSn}

\bibitem{NaH4NO3}
\emph{NaH$_{4}$NO$_{3}$ Structures}. \url{https://aflow.org/p/5YKM/} and links
  therein.
\input{NaH4NO3}

\bibitem{OpenKIM}
E.~B. Tadmor, R.~S. Elliot, J.~P. Sethna, R.~E. Miller, and C.~A. Becker,
  \emph{The potential of atomistic simulations and the knowledgebase of
  interatomic models}, JOM \textbf{63}, 17 (2011),
  \doi{10.1007/s11837-011-0102-6}.
\input{OpenKIM}

\bibitem{kim-tools}
\emph{kim-tools}. \url{https://github.com/openkim/kim-tools}.
\input{kim-tools}

\bibitem{NOMAD_2023}
M.~Scheidgen, L.~Himanen, A.~N. Ladines, D.~Sikter, M.~Nakhaee, \'{A}d\'{a}m
  Fekete, T.~Chang, A.~Golparvar, J.~A. M\'{a}rquez, S.~Brockhauser,
  S.~Br\"{u}ckner, L.~M. Ghiringhelli, F.~Dietrich, D.~Lehmberg, T.~Denell,
  A.~Albino, H.~Näsström, S.~Shabih, F.~Dobener, M.~K\"{u}hbach, R.~Mozumder,
  J.~F. Rudzinski, N.~Daelman, J.~M. Pizarro, M.~Kuban, C.~Salazar,
  P.~Ondra\v{c}ka, H.-J. Bungartz, and C.~Draxl, \emph{NOMAD: A distributed
  web-based platform for managing materials science research data}, JOSS
  \textbf{8}, 5388 (2023), \doi{10.21105/joss.05388}.
\input{NOMAD_2023}

\bibitem{curtarolo:art208_UPDATED}
M.~L. Evans, J.~Bergsma, A.~Merkys, C.~W. Andersen, O.~B. Andersson,
  D.~Beltr\'{a}n, E.~Blokhin, T.~M. B. R.~C. Balderas, K.~Choudhary, A.~D.
  D\'{i}az, R.~D. Garc\'{i}a, H.~Eckert, K.~Eimre, M.~E.~F. Montero, A.~M.
  Krajewski, J.~J. Mortensen, J.~M. N\'{a}poles-Duarte, J.~Pietryga, J.~Qi,
  F.~T. Carrillo, A.~Vaitkus, J.~Yu, A.~C. Zettel, P.~B. de~Castro,
  J.~Carlsson, T.~F.~T. Cerqueira, S.~Divilov, H.~Hajiyani, F.~Hanke, K.~Jose,
  C.~Oses, J.~Riebesell, J.~Schmidt, D.~Winston, C.~Xie, X.~Yang, S.~Bonella,
  S.~Botti, S.~Curtarolo, C.~Draxl, L.~E.~F. Cobas, A.~Hospital, Z.-K. Liu,
  M.~A.~L. Marques, N.~Marzari, A.~J. Morris, S.~P. Ong, M.~Orozco, K.~A.
  Persson, K.~S. Thygesen, C.~Wolverton, M.~Scheidgen, C.~Toher, G.~J. Conduit,
  G.~Pizzi, S.~Gra{\v{z}}ulis, G.-M. Rignanese, and R.~Armiento,
  \emph{{Developments and applications of the OPTIMADE API for materials
  discovery, design, and data exchange}}, Digit. Discov. \textbf{3}, 1509--1533
  (2024), \doi{10.1039/D4DD00039K}.
\input{curtarolo:art208_UPDATED}

\bibitem{pymatgen}
\emph{pymatgen}. \url{https://pymatgen.org/}.
\input{pymatgen}

\bibitem{Robocrystallographer}
A.~M. Ganuse and A.~Jain, \emph{Robocrystallographer: automated crystal
  structure text descriptions and analysis}, MRS\ Commun. \textbf{9}, 874--881
  (2019), \doi{10.1557/mrc.2019.94}.
\input{Robocrystallographer}

\bibitem{robocrystallographer_source}
\emph{robocrystallographer}.
  \url{https://github.com/hackingmaterials/robocrystallographer}.
\input{robocrystallographer_source}

\bibitem{Jmol}
\emph{{Jmol}: an open-source {J}ava viewer for chemical structures in 3{D}},
  http://www.jmol.org/.
\input{Jmol}

\bibitem{Jmol_Hanson}
R.~M. Hanson, \emph{\textit{Jmol} {---} a paradigm shift in crystallographic
  visualization}, J.\ Appl.\ Crystallogr. \textbf{43}, 1250--1260 (2010),
  \doi{10.1107/S0021889810030256}.
\input{Jmol_Hanson}

\bibitem{aviary}
\emph{aviary}. \url{https://github.com/CompRhys/aviary}.
\input{aviary}

\bibitem{riebesell_2025_framework}
J.~Riebesell, R.~E. Goodall, P.~Benner, Y.~Chiang, B.~Deng, G.~Ceder, M.~Asta,
  A.~A. Lee, A.~Jain, and K.~A. Persson, \emph{A framework to evaluate machine
  learning crystal stability predictions}, Nat. Mach. Intell. \textbf{7},
  836--847 (2025).
\input{riebesell_2025_framework}

\bibitem{WyckoffDiff_source}
\emph{WyckoffDiff}. \url{https://github.com/httk/WyckoffDiff}.
\input{WyckoffDiff_source}

\bibitem{pmlr-v267-ekstrom-kelvinius25a}
F.~Ekstr\"{o}m~Kelvinius, O.~B. Andersson, A.~S. Parackal, D.~Qian,
  R.~Armiento, and F.~Lindsten, \emph{{W}yckoff{D}iff – A Generative
  Diffusion Model for Crystal Symmetry}, in \emph{Proceedings of the 42nd
  International Conference on Machine Learning}, edited by A.~Singh, M.~Fazel,
  D.~Hsu, S.~Lacoste-Julien, F.~Berkenkamp, T.~Maharaj, K.~Wagstaff, and J.~Zhu
  (PMLR, 2025), vol. 267, pp. 15130--15147.
\input{pmlr-v267-ekstrom-kelvinius25a}

\bibitem{Kazeev202529495}
N.~Kazeev, W.~Nong, I.~Romanov, R.~Zhu, A.~Ustyuzhanin, S.~Yamazaki, and
  K.~Hippalgaonkar, \emph{Wyckoff Transformer: Generation of Symmetric
  Crystals} (2025), vol. 267, p. 29495 – 29526.
\input{Kazeev202529495}

\bibitem{Wyckoff_Transformer_source}
\emph{Wyckoff Transformer}.
  \url{https://github.com/SymmetryAdvantage/WyckoffTransformer}.
\input{Wyckoff_Transformer_source}

\bibitem{ARISE_2021}
L.~Andreas, Z.~Angelo, and L.~M. Ghiringhelli, \emph{Robust recognition and
  exploratory analysis of crystal structures via Bayesian deep learning}, Nat.\
  Commun. \textbf{12}, 6234 (2021), \doi{10.1038/s41467-021-26511-5}.
\input{ARISE_2021}

\bibitem{ARISE_source}
\emph{ARISE}. \url{https://github.com/angeloziletti/ai4materials}.
\input{ARISE_source}

\bibitem{simmate}
\emph{simmate}. \url{https://simmate.org}.
\input{simmate}

\bibitem{Wilkinson_FAIR_SciData_2016_UPDATED}
M.~D. Wilkinson, M.~Dumontier, I.~J. Aalbersberg, G.~Appleton, M.~Axton,
  A.~Baak, N.~Blomberg, J.~W. Boiten, L.~B. {da Silva Santos}, P.~E. Bourne,
  J.~Bouwman, A.~J. Brookes, T.~Clark, M.~Crosas, I.~Dillo, O.~Dumon,
  S.~Edmunds, C.~T. Evelo, R.~Finkers, A.~Gonzalez-Beltran, A.~J.~G. Gray,
  P.~Groth, C.~Goble, J.~S. Grethe, J.~Heringa, P.~A.~C. {'t Hoen}, R.~Hooft,
  T.~Kuhn, R.~Kok, J.~Kok, S.~J. Lusher, M.~E. Martone, A.~Mons, A.~L. Packer,
  B.~Persson, P.~Rocca-Serra, M.~Roos, R.~{van Schaik}, S.-A. Sansone,
  E.~Schultes, T.~Sengstag, T.~Slater, G.~Strawn, M.~A. Swertz, M.~Thompson,
  J.~{van der Lei}, E.~{van Mulligen}, J.~Velterop, A.~Waagmeester,
  P.~Wittenburg, K.~Wolstencroft, J.~Zhao, and B.~Mons, \emph{The FAIR Guiding
  Principles for scientific data management and stewardship}, Sci.\ Data
  \textbf{3}, 160018 (2016), \doi{10.1038/sdata.2016.18}.
\input{Wilkinson_FAIR_SciData_2016_UPDATED}

\bibitem{NOMAD_about}
\emph{NOMAD}. \url{https://nomad-lab.eu/nomad-lab/} (Accessed 22 Jul 2026).
\input{NOMAD_about}

\bibitem{OPTIMADE_about}
M.~Consortia, \emph{OPTIMADE list of providers}.
  \url{https://www.optimade.org/providers-dashboard/} (Accessed 23 Jul 2026).
\input{OPTIMADE_about}

\bibitem{aflowXTALFINDER}
D.~Hicks, C.~Toher, D.~C. Ford, F.~Rose, C.~{De Santo}, O.~Levy, M.~J. Mehl,
  and S.~Curtarolo, \emph{{AFLOW-XtalFinder}: a reliable choice to identify
  crystalline prototypes}, npj\ Comput.\ Mater. \textbf{7}, 30 (2021),
  \doi{10.1038/s41524-020-00483-4}.
\input{aflowXTALFINDER}

\bibitem{CMS_Ong2012b}
S.~P. Ong, W.~D. Richards, A.~Jain, G.~Hautier, M.~Kocher, S.~Cholia,
  D.~Gunter, V.~L. Chevrier, K.~A. Persson, and G.~Ceder, \emph{{Python
  Materials Genomics (pymatgen): A robust, open-source python library for
  materials analysis}}, Comput.\ Mater.\ Sci. \textbf{68}, 314--319 (2013),
  \doi{10.1016/j.commatsci.2012.10.028}.
\input{CMS_Ong2012b}

\bibitem{goodall_2022_rapid}
R.~E. Goodall, A.~S. Parackal, F.~A. Faber, R.~Armiento, and A.~A. Lee,
  \emph{Rapid discovery of stable materials by coordinate-free coarse
  graining}, Sci.\ Adv. \textbf{8}, eabn4117 (2022).
\input{goodall_2022_rapid}

\end{thebibliography}
\end{document}